\documentclass[10pt,letterpaper]{article}
\usepackage{graphicx}
\usepackage{float}
\usepackage[top=0.85in,left=2.75in,footskip=0.75in]{geometry}

\usepackage{amsmath,amssymb}

\usepackage{changepage}

\usepackage[utf8x]{inputenc}

\usepackage{textcomp,marvosym}

\usepackage{cite}

\usepackage{nameref,hyperref}

\usepackage[right]{lineno}

\usepackage{microtype}
\DisableLigatures[f]{encoding = *, family = * }

\usepackage[table]{xcolor}

\usepackage{array}

\newcolumntype{+}{!{\vrule width 2pt}}

\newlength\savedwidth

\raggedright
\usepackage[aboveskip=1pt,labelfont=bf,labelsep=period,justification=raggedright,singlelinecheck=off]{caption}

\makeatletter
\renewcommand{\@biblabel}[1]{\quad#1.}
\makeatother

\usepackage{lastpage,fancyhdr,graphicx}
\usepackage{epstopdf}
\fancyheadoffset[L]{2.25in}
\fancyfootoffset[L]{2.25in}
\begin{document}
\vspace*{0.2in}

\begin{flushleft}
{\Large
\textbf\newline{Equilibrium properties of $\emph{E.\  coli}$ lactose permease cotransport—a random-walk model approach} 
}
\newline
\\
Haoran Sun\textsuperscript{1*},
\\
\bigskip
\textbf{1} School of Mathematical Sciences, Fudan University, Shanghai, China
\\
\bigskip

%
%





* 18300180051@fudan.edu.cn

\end{flushleft}
\section*{Abstract}
The symport of lactose and $\rm H^+$ is an important physiological process in $E.\ coli$, for it is closely related to cellular energy supply. In this paper, the symport of $\rm H^+$ and lactose by $E.\ coli$ LacY protein is computationally simulated using a newly proposed cotransport model that takes the ``leakage" phenomenon (uncoupled sugar translocation) into account and also satisfies the static head equilibrium condition. Then, we study the equilibrium properties of this cotransport process including equilibrium solution and the time required to reach equilibrium state, when varying the parameters of the initial state of cotransport system. It can be found that in the presence of leakage, $\rm H^+$ and lactose will reach their equilibrium state separately, but the intensity of ``leakage" has almost no effect on the equilibrium solution, while the stronger leakage is, the shorter the time required for $\rm H^+$ and lactose to reach equilibrium. For $E.\ coli$ cells with different periplasm and cytoplasm volumes, when cotransport is performed at constant initial particle concentration, the time for cytoplasm pH to be stabilized increases monotonically with the periplasm to cytoplasm volume ratio. For a certain $E.\ coli$ cell, as it continues to lose water and contract, the time for cytoplasm pH to be stabilized by cotransport also increases monotonically when the cell survives. The above phenomena and other findings in this paper may help us to not only further validate or improve the model, but also deepen our understanding of the cotransport process of $E.\ coli$ LacY protein.


\section*{Introduction}
	LacY protein (lactose permease) of $Escherichia\ coli$ is a kind of transport protein involved in the secondary active transport of hydrogen ions and lactose molecules. The protein uses the energy stored in the $\rm H^+$ electrochemical potential to cotransport hydrogen ions into the cytoplasm along with $\rm\beta-galactosides$, such as lactose\cite{guan2006lessons}. Since large amounts of lactose are required to sustain life activities in $E.\ coli$ cells, the intracellular lactose concentration is usually higher than the environment, and the above cotransport process is usually inverse to the lactose concentration gradient\cite{abramson2003structure}. Possible mechanisms and mathematical models for the cotransport process have been extensively studied, starting roughly from Cohen and Rickenberg's report in 1955\cite{kramer2014membrane}. Among the numerous research results so far, one of the most significant achievement, and also a mechanism now generally accepted, is the ``six-state" mechanism described by Kaback $et\ al.$ in their 2001 article\cite{kaback2001kamikaze}, which is baesd on a more universal cotransport mechanism proposed by Jardetzky in 1966\cite{jardetzky1966simple}. The so-called ``six-state" mechanism refers to the fact that the cotransporter lactose permease have six different functional conformations (states) in the cotransport process. The LacY protein begins the reaction cycle at a outward-facing state 1, quickly binds a hydrogen ion, turning to state 2, then continues to trap a lactose molecule and turns to state 3, during which the cotransporter remains in the outward-facing state. After the combination of particles, the cotransporter makes a rapid conformational change to inward-facing, followed by detachment of the lactose molecule to state 5, then to state 6 by shedding the hydrogen ion, and finally returns to state 1 by another rapid conformational change. All above processes are naturally reversible. However, the former conventional mechanism is challenged by some subsequent experimental results\cite{andrini2008leak}, where uncoupled transport is observed and determined, $i.e.$ , the two kinds of particles involved in cotransport do not follow the previous stoichiometric, and one kind has a uniport-like``leakage" phenomenon. It also poses a problem for how to modify previous existing mathematical models that simulate this ``six-state" mechanism. For the symport of sodium ions with glucose by SGLT1 protein, which is similar to the cotransport process of $E.\ coli$ LacY protein, several possible mechanisms leading to the leakage phenomenon have been proposed in the literature\cite{centelles1991energetic}, among which the simpler one that has also been used many times to modify mathematical models, is to allow the transition between cotransporter states 2 and 5. Nevertheless, some literatures such as\cite{naftalin2010reassessment} propose that any determinate modified cotransport model by allowing the transition between states 2 and 5 cannot satisfy a certain thermodynamic condition (static head equilibrium condition). Then Barreto $et\ al.$ propose a statistical mechanical model based on the above way of modifying conventional mechanism on the work of paper\cite{barreto2019transport}, which is a random-walk model and satisfies the static head equilibrium condition in the case of leakage\cite{barreto2020random}.

	Based on such a model, we are finally able to computationally simulate the transport process of $E.\ coli$ LacY protein in the presence of leakage, and find some distinct equilibrium state properties in the presence and absence of leakage, such as the equilibrium solution and the convergence rate to the equilibrium solution. In this paper, we first briefly restate and analyze the model in article\cite{barreto2020random} and predict the results of the computational simulations in the following. Next, we vary the intensity of the leakage, the volume of $E.\ coli$ periplasm and cytoplasm, the initial concentration of the particles and the number (density) of cotransporter to study the variation of the corresponding equilibrium solution and the time required to reach equilibrium. Finally we find some very similar or different phenomena in the presence and absence of leakage.

\section*{A brief statement and analysis of the model}
\begin{figure}[!h]
\centering
\includegraphics[scale=0.7]{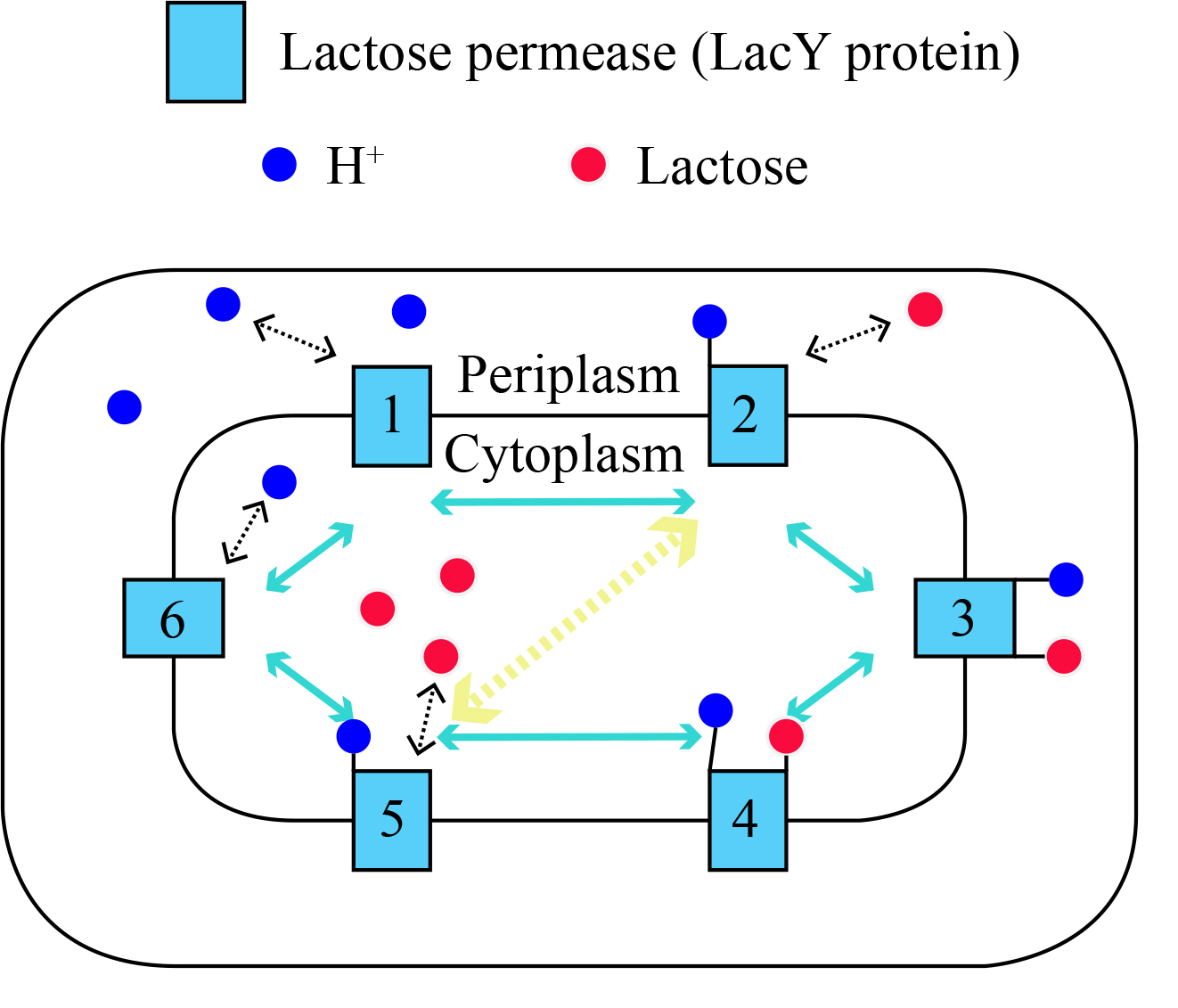}
\caption{{\bf The modified six-state mechanism of cotransport.}
An alternating access model of $E.\ coli$ $\rm lactose-H^+$ cotransport by Jardetzky modified with the transition between cotransporter states 2 and 5. A simple schematic of the ``six-state" mechanism of cotransporter LacY protein of $E.\ coli$ during cotransport is shown in the figure above. Numbers 1 to 6 indicate state 1 to state 6, which are the six effective conformations of LacY protein. The blue double-head arrows refer to the six states of cotransporter in relation to each other, and the yellow dashed arrow refers to the way to modify the present model—allowing the transition between cotransporter states 2 and 5.}
\label{1}
\end{figure}

	The model in article\cite{barreto2020random} considers a closed system of periplasm and cytoplasm, the volumes of which are $V_p$ and $V_c$ respectively. Periplasm and cytoplasm are separated by a cell membrane with cotransporters embedded, across which there is a membrane potential $\Delta\Psi$ that we assume a constant (cytoplasm lower). The total number of cotransporters on the cell membrane is fixed to $n$, and they are all assumed to be independent of each other. Let $n_k,k=1,2\cdots6,$ represent the number of cotransporters in the current state $k$, respectively, and the sum of these six terms is clearly the fixed value $n$. Here I use an illustration  Fig~\ref{1} to graphically show the six-state mechanism described in the introduction section for the readers to understand it more visually.

	$N_c^{\rm H^+}(t)$ and $N_c^{\rm L}(t)$ represent the current number of hydrogen ions and lactose molecules in cytoplasm, and $N_p^{\rm H^+}(t)$ and $N_p^{\rm L}(t)$ similarly for periplasm. When it is not necessary to specify one kind of particle or one of periplasm and cytoplasm, we will later use $S$ to refer to either of the two kinds of particles involved in cotransport and $l$ refer to periplasm or cytoplasm. Since the system is closed, the total number of $S$ particles in all states and positions is always equal to the number we put in at the beginning, as a constant value, denoted $N_S$. $\xi$ represents the intensity of leakage (similar to the ratio of leakage current to cotransport current excluding leakage) and takes the value in $[0,1]$. $\xi=0$ means no leakage occurs, $\xi=1$ means the cotransporter in state 2 has an equal probability to take a conformation change to state 1,3 or 5, and the larger the value of $\xi$, the more obvious the leakage phenomenon is. The master equations of the model consists of ten differential equations related to each other, the exact form of which is not repeated here; please move to the original \cite{barreto2020random}. Here we perform a brief analysis of the model.

	The original \cite{barreto2020random} first makes the left-hand side quotient of the master equation tend to 0 by making $t\rightarrow\infty$, which causes all variables to take equilibrium solutions (or asymptotic solutions), and the solutions when $\xi=0$ and $\xi\neq0$ have the following form,
\begin{displaymath}
N_{l,\xi=0}^S=C_0V_l\left(\frac{\alpha_{K+1K}}{\alpha_{KK+1}}\frac{n_{K+1}}{n_K}\right)^{h/\nu_S} ,
\end{displaymath}	
\begin{displaymath}
N_{l,\xi\neq0}^S=\left(1-\frac{\xi}{3}\right)^{\tilde{h}/\nu_S}\!\!\!\!\!\!\!\!\!C_0V_l\left(\frac{\alpha_{K+1K}}{\alpha_{KK+1}}\frac{n_{K+1}}{n_K}\right)^{h/\nu_S} .
\end{displaymath}
	$\nu_S$ is the number of particle $S$ transported in one cycle of cotransporter in figure\ref{1}, here equals 1 for $\rm H^+$ and lactose. $h, K, \tilde{h}, \alpha_{pq}$ are all constants decided by $S$ and $l$, and the details are shown in article\cite{barreto2020random}. After obtaining the above results, the paper \cite{barreto2020random} states that $N_{l,\xi\neq0}^S=\left(1-\frac{\xi}{3}\right)^{\tilde{h}/\nu_S}N_{l,\xi=0}^S$, which leads to the equilibrium solution for all values of $\xi$. However, a careful observation shows that the above two equations are essentially the relationship between the equilibrium solution of the particle numbers $N_{l,\xi}^S$ and the cotransporter numbers $n_K$. The equilibrium solutions of the cotransporters are not necessarily the same for different values of $\xi$ (in practice, it is clear from the calculations below that they are indeed not the same), so the conclusion of the original equilibrium solution in \cite{barreto2020random} does not hold.

	In fact, we can get new information about the equilibrium solution by the Gibbs free energy change as well as the electrochemical gradient. The original\cite{barreto2020random} proves that the equilibrium solution satisfies the following static head equilibrium condition when $\xi=0$, 
\begin{displaymath}
\frac{N_c^A/V_c}{N_p^A/V_p}=\left(\frac{N_p^B/V_p}{N_c^B/V_c}\right)^{(2f-1)\nu_B/\nu_A}\!\!\!\!\!\!\!\!\!\!\!\!\!\!\!\!\!\!\!\!\!\!\!\times\exp\left(-\frac{1}{k_BT}\left[Q_A+(2f-1)\frac{\nu_A}{\nu_B}Q_B\right]\Delta\Psi\right).
\end{displaymath}
	Here $f=1$ for lactose permease and $A,B$ are particles transported by cotransporter. And the following scaling relationships between equilibrium solutions are obtained in the same way when $\xi\neq0$ (also satisfying the static head equilibrium),
\begin{displaymath}
\frac{N_c^S/V_c}{N_p^S/V_p}=\exp\left(-\frac{1}{k_BT}Q_S\Delta\Psi\right) .
\end{displaymath}
	$Q_S$ is the charge of a particle $S$, $k_B$ is the Boltzmann constant, and $T$ is the is the ambient temperature of $E.\ coli$. The above two relations are important constraints on the equilibrium solution, since in reality the number of cotransporters is much smaller than the number of particles involved in cotransport, and the total number of particles is constant during the transport, $i.e.$ the following two equations hold at all times during the transport, 
\begin{displaymath}
N_A-N_c^A-N_p^A=\nu_A[n_3+n_4+(n_2+n_5)f],
\end{displaymath}
\begin{displaymath}
N_B-N_c^B-N_p^B=\nu_B[(n_3+n_4)f+(1-f)(n_1+n_6)].
\end{displaymath}
	Then we can see that under the general condition that the total number of cotransporters is very small, the equilibrium solution changes quite little with the parameter $\xi$ when $\xi\neq0$. Only if the total number of cotransporters is not small relative to the total number of particles and also the distribution of cotransporters in each state varies considerably when $\xi$ changes does the equilibrium solution change significantly. The computational verification of the above theoretical descriptions will be carried out below, while the above conclusions give directional hints for our later calculations.
\section*{Computational simulations and equilibrium property studies using the above model}
	In the following of this paper, computational simulations of the $E.\ coli$ LacY protein 1:1 symport processes of $\rm H^+$ and lactose are performed with the aforementioned model. The values used in the calculation are from the article \cite{barreto2020random}\cite{barreto2019transport} or selected according to the reality, as detailed in the calculation section later. We use the finite difference method to solve the differential master equations of the model numerically\cite{bathe2006finite}, that is, we use the relation ${dN(t)}/{dt}\approx{(N(t+\Delta t)-N(t))}/{\Delta t}$ to transform the differential equation into a difference equation for calculation, and in subsequent calculations in this paper we take the time step $\Delta t=0.1$
	\subsection*{Effect of parameter $\xi$ (leakage intensity) on equilibrium properties}
	The $\xi=0$ and $\xi\neq0$ cases are too different to be studied together, so we focus on the $\xi\neq0$ case. First, the correlation between the equilibrium solution and $\xi$ will be verified. The following Fig~\ref{2} shows the correlation between $\xi$ and the equilibrium solutions as well as that between $\xi$ and the time to reach equilibrium states for $\rm H^+$ and lactose at a fixed initial condition, respectively.

\begin{figure}[!h]
\centering
\includegraphics[scale=0.8]{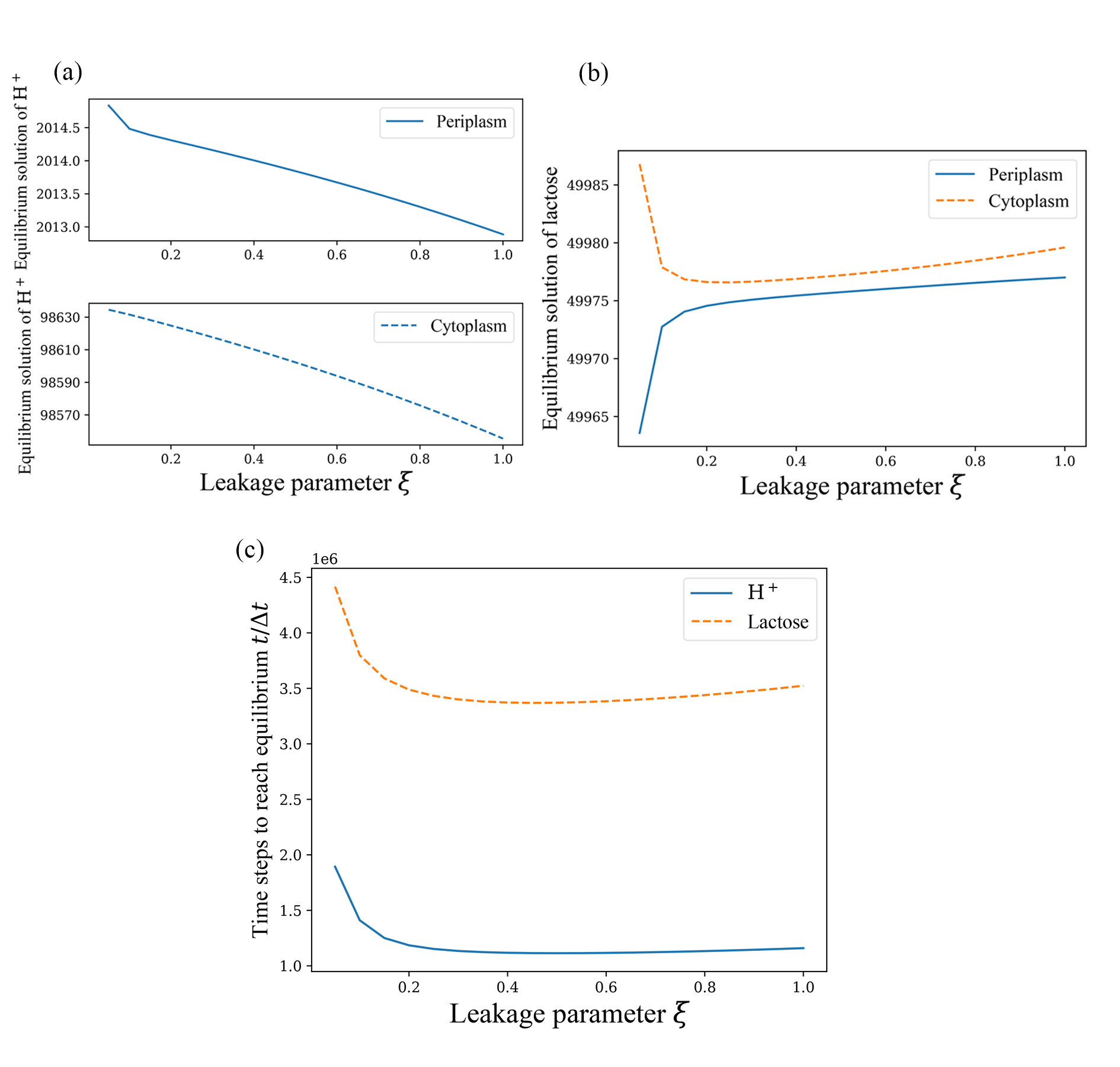}
\caption{{\bf Equilibrium properties of lactose permease cotransport in $E.\ coli$ as $\xi$ changes.}
Only the $\xi\neq0$ cases are shown with the following initial conditions: $n_1=10^3, n_{k\neq1}=0, N_p^{\rm H^+}=10^5, N_p^{\rm L}=2.5\times10^4, N_c^{\rm L}=7.5\times10^4, N_c^{\rm H^+}=10^3$. And also, we set $V_c=V_p=10^6/C_0$, $C_0$ is a constant with the unit of $\rm nm^{-3}$. The meaning of $C_0$ is the same as which in \cite{barreto2020random}, similarly hereinafter. (a) The equilibium solution of $\rm H^+$ in periplasm and cytoplasm. (b) The equilibium solution of lactose in periplasm and cytoplasm. (c) Time used to reach equilbrium state for $\rm H^+$ and lactose with the unit of $\Delta t$.}
\label{2}
\end{figure}

	The reason why the initial conditions being so is that, according to the article \cite{wilks2007ph}, $E.\ coli$ cells are generally in an environment that causes the pH of periplasm about 1.8 to 2.1 higher than that of cytoplasm. From the above Fig~\ref{2}(a) and Fig~\ref{2}(b), it can be found that in the case that $\xi$ is not very small ($\geq0.1$), the above general conclusions about the equilibrium solution (insignificant variation of the equilibrium solution with $\xi$ and general proportionality between the equilibrium solutions) are correct within error permissibility. In fact, when the total number of cotransporters is increased (in this example we just need to increase $n_1$), the error will be significantly reduced and the proportional relationship between the equilibrium solutions will become more obvious and precise.

	Another important feature about cotransport is the time required to reach the equilibrium solution. Although the equilibrium solution is asymptotic and theoretically not achievable in finite time, we can still consider that equilibrium has been reached when the change rate or the first-order backward difference quotient of the solution is small enough in our simulation. When there is no leakage ($\xi=0$), it is easy to find that $\rm H^+$ and lactose reach equilibrium at the same time by calculation, which is also easy to see from the above theoretical analysis. However, when leakage does exist ($\xi\neq0$), the time for the two kinds of particles to reach equilibrium is seperated, which can also be inferred by splitting the full reaction cycle into two independent transport process. The above Fig~\ref{2}(c) shows the relationship between the time of two kinds of particles to reach equilibrium and $\xi$ when $\xi\in(0,1]$.

	Fig~\ref{2}(c) is calculated by considering that when $[N_c^S(t)-N_c^S(t-\Delta t)]/N_c^S(t)\leq10^{-9}$ holds for almost every discrete time moments from some t (for over $99.9999\%$ of time moments), the particle $S$ reach the equilibrium. Subsequent calculations of the time to reach the equilibrium are similar, but accuracy may be adjusted as needed. The reasons for not using the first-order backward difference quotient and instead using change rate are the poor estimation of the range of the difference quotient and the fact that the difference quotient in this model is not guaranteed to be consistently smaller than the required accuracy after a certain time. As can be seen in the figure, the time for $\rm H^+$ to reach equilibrium is always shorter than the time for lactose when $\xi\neq0$. Because we can formally split the overall transort process in the presence of leakage, the full reaction process can be viewed as a combination of a leakage-free cotransport process and a separate uniport process of $\rm H^+$, with little interference between them. Then $\rm H^+$ has two transport processes transporting it, while lactose has only one. So $\rm H^+$ can reach equilibrium with the net transport of the two processes canceling each other, while lactose has not reached equilibrium at that point. It can also be observed that the time for the two kinds of particles to reach equilibrium do not have a consistent trend with the change of $\xi$. Both of them are large when $\xi$ is small, after that it decreases rapidly with the increase of $\xi$, and the change is not obvious after $\xi>0.2$. But the equilibrium time of lactose has a small rebound after $\xi>0.5$, while $\rm H^+$ continues to remain monotonically decreasing. The reason for this phenomenon is not clear, and it is speculated that there may be some calculation errors. Another point worth mentioning is that particle numbers converge much faster when $\xi=0$ than $\xi\neq0$. The addition of the transition between cotransporter states 2 and 5 ($i.e.$ , making $\xi\neq0$) makes the convergence of the system much slower.
\subsection*{Effect of periplasm and cytoplasm volumes $V_p,V_c$ on equilibrium properties}
	The previous arithmetic examples have been calculated assuming $V_p=V_c$, and as seen before, the proportionality satisfied by the equilibrium solution when $\xi\neq0$ is actually a proportionality between the particle concentrations in the two reaction chambers (periplasm and cytoplasm), so we consider changing the periplasm and cytoplasm volume ratios to verify the conclusion and observe if there are new phenomena. Also due to the previous conclusions on the equilibrium solution and the convergence rate in the case of $\xi\neq0$, we can only study the $\xi=0$ and $\xi=1$ cases. The equilibrium solutions of the $\xi=0$ case are shown in the following Fig~\ref{3}. 
\begin{figure}[!h]
\centering
\includegraphics[scale=0.7]{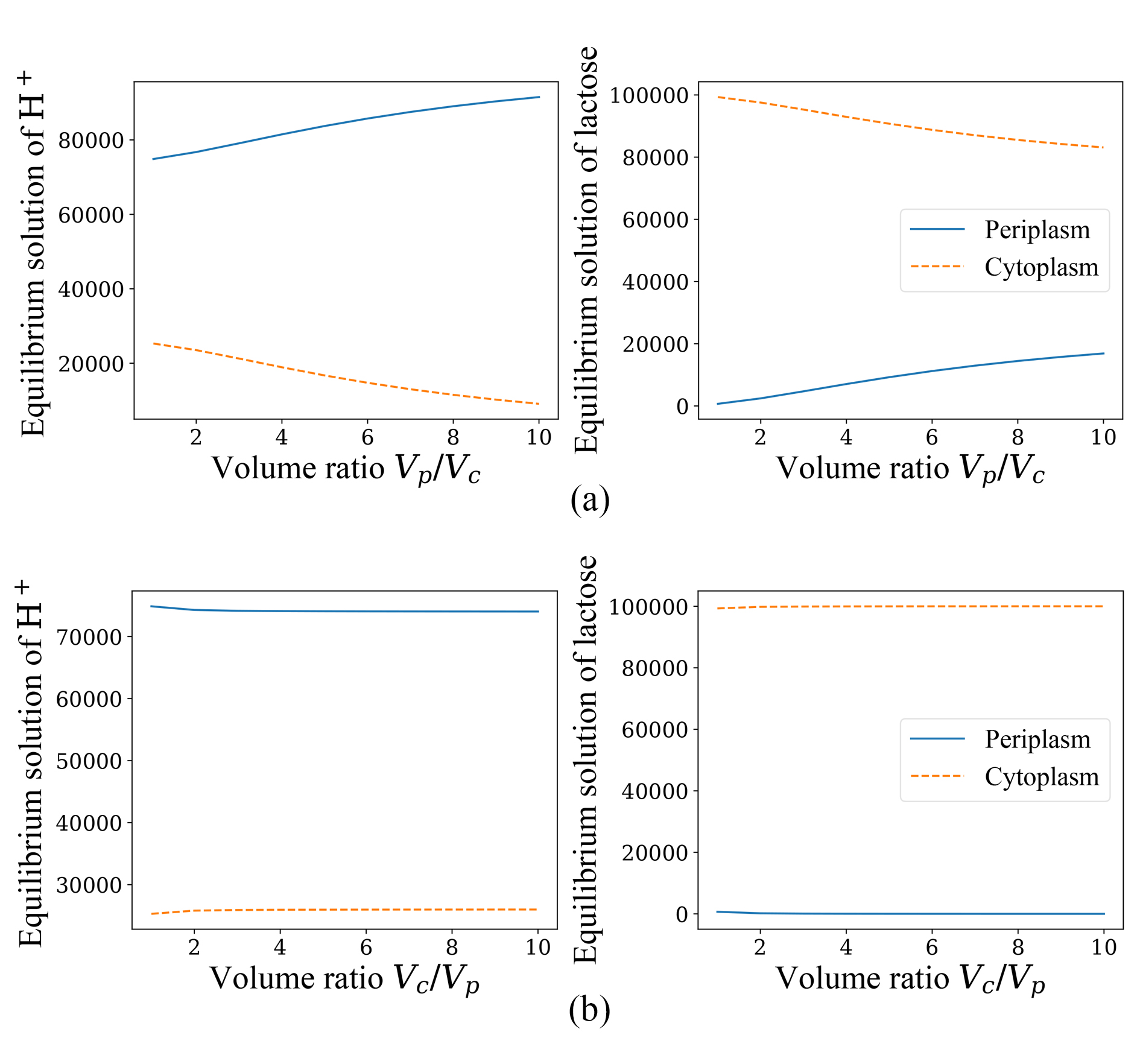}
\caption{{\bf The equilibium solutions for $\xi=0$ when $V_c$ or $V_p$ is fixed and the other one varies.}
(a) The equilibium solutions of $\rm H^+$ and lactose when $V_c$ is fixed to $10^6/C_0$ and $V_p/V_c>1$, with initial conditions $n_1=10^3, n_{k\neq1}=0, N_p^{\rm H^+} =10^5$, $N_p^{\rm L}=2.5\times10^4$, $N_c^{\rm L}=7.5\times10^4, N_c^{\rm H^+}=10^3$. (b) The equilibium solutions of $\rm H^+$ and lactose when $V_c$ is fixed to $10^6/C_0$ and $V_p/V_c<1$, with the same initial conditions before.}
\label{3}
\end{figure}

	From Fig~\ref{3}, the following phenomenon can be found that under the condition of fixing the number of particles involved in cotransport ($i.e.$ fixing $N_c^S,N_p^S$ in initial conditions), the change of $V_c$ seems to have less effect on the equilibrium solution than $V_p$, and the system equilibrium solution is more insensitive to the change of $V_c$. When changing other initial conditions and making $V_p$ a constant value to change $V_c/V_p$, it is found that this phnomenon is actually more accurately formulated as that changing $V_c$ or $V_p$ only under the condition that $V_p/V_c>1$ has a more pronounced effect on the equilibrium solution when $\xi=0$, thus showing an insensitivity to the increase in the denominator $V_c$. After adjusting the parameters and performing a lot of calculations, it is found that this property is somewhat related to the membrane potential. In this case the direction of potential decrease is from periplasm to cytoplasm, and the membrane potential is $-100mV$. However, when decreasing this potential difference, this phenomenon becomes less and less obvious, especially when reversing the membrane potential, $i.e.$ making the potential of cytoplasm higher than that of periplasm, the significances of $V_c$ and $V_p$ in the above properties are switched, and $V_p$ becomes the one that has less influence on the equilibrium solution of the system. 

	The above phenomena can be explained from the theoretical analysis, for which we consider the Gibbs free energy variation at the starting moment $\Delta G$. For this cotransport process of the $E.\ coli$ LacY protein, we can rewrite the free energy variation in terms of the sum of the chemical potential multiplying the stoichiometry of the two particles \cite{nelson2008lehninger}, bringing the data and slightly transforming then we have, 
\begin{displaymath}
\Delta G=k_BT\ln\left(\frac{N_c^{\rm H^+}N_c^{\rm L}}{N_p^{\rm H^+}N_p^{\rm L}}\right)+Q_{\rm H^+}\Delta\Psi+2k_BT\ln(V_p/V_c).
\end{displaymath}
	For a fixed initial condition, the first of the three terms on the right-hand side of the above equation is constant, and based on the data we use in calculation and the data corresponding to the environment in which $E.\ coli$ is usually found, this term is usually in the $(-4k_BT,k_BT)$ interval, and the second term is usually around $-3.9k_BT$ \cite{lo2007nonequivalence}. When changing $V_p$ or $V_c$, the absolute value of $\Delta G$ can be reduced only if $V_p/V_c>1$, and making $\left|\Delta G\right|$ decrease produces a larger rate of change than making it increase when the amount of change in $V_p/V_c$ is the same. The equilibrium solution is reached when $\Delta G$ is 0, and it is easy to find that $\Delta G$ varies monotonically during the cotransport process. So $\Delta G$ can actually represent the "gap" between the initial state and the equilibrium state, and the gap between the equilibrium solutions is larger when the difference between $\Delta G$ is larger. And since the negative membrane potential in this process contributes positively (also not small proportion) to $\Delta G$, it produces a very different property when the membrane potential decreases or even reverses.

	Next, according to common biological sense, we will observe the relationship between $V_p/V_c$ or $V_c/V_p$ and the time required for the system to reach equilibrium when $V_p/V_c<1$. In the following, the case of $\xi=0$ comes first. Inspired by the above phenomena, we fix $V_c$ and $V_p$ separately and change the other one to calculate. For practical reasons, particle numbers should not be fixed as in the case of equilibrium solution just probed before, but the concentration of particles in periplasm and cytoplasm in initial conditions should be fixed, so that the results are more realistic. The following Fig~\ref{4}(a) and Fig~\ref{4}(b) are plotted with $V_p/V_c$ as the horizontal coordinate (with $V_c$ fixed) and $V_c/V_p$ as the horizontal coordinate (with $V_p$ fixed), respectively, and the time to reach the equilibrium solution as the vertical coordinate.
\begin{figure}[!h]
\centering
\includegraphics[scale=0.9]{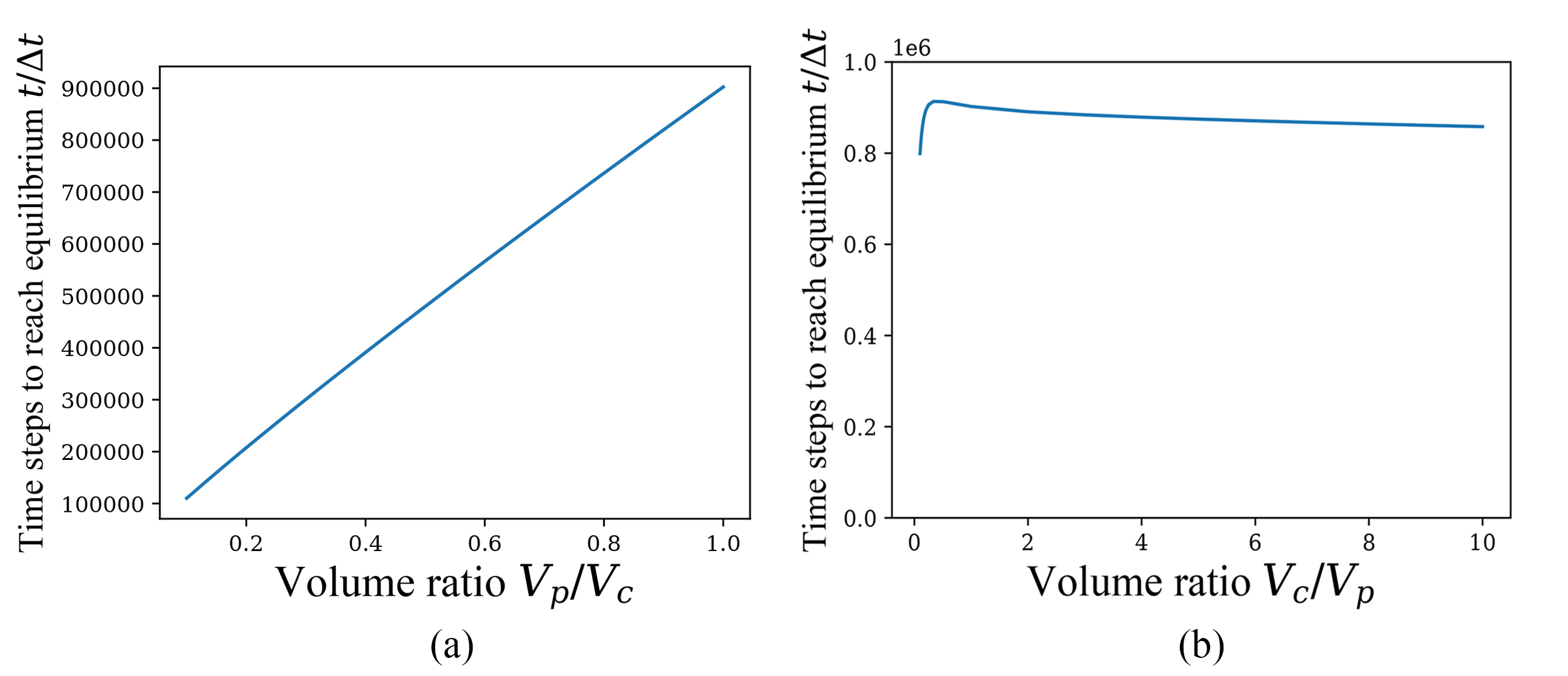}
\caption{{\bf Time to reach equilibrium for $\xi=0$ when $V_c$ or $V_p$ is fixed and the other one varies.}
Time used to reach equilbrium state with the unit of $\Delta t$ when the initial concentrations of $\rm H^+$ and lactose are fixed in periplasm and cytoplasm, which has the initial conditions $n_1=10^3, n_{k\neq1}=0, N_p^{\rm H^+}/V_p=C_0/10, N_c^{\rm H^+}/V_c=C_0/1000, N_p^{\rm L}/V_p=C_0/40, N_c^{\rm L}/V_c=3C_0/40$, $\xi=0$. (a) $V_c$ fixed to $10^6/C_0$. (b) $V_p$ fixed to $10^6/C_0$.}
\label{4}
\end{figure}

	When the change rate of $N_c^S$ is always less than $10^{-9}$, the particle $S$ is considered to reach equilibrium in the above Fig~\ref{4}. It is still clear to find different effects of $V_p$ and $V_c$ on the time required to reach equilibrium, and the effect of $V_c$ is still much smaller than that of $V_p$. Increasing $V_p$ when $V_c$ is fixed can monotonically increase the time for the system to reach equilibrium. However, changing $V_c$ when $V_p$ is fixed has a less significant effect on the convergence rate. This phenomenon can still be explained using the free energy variation $\Delta G$, because although $\Delta G$ is constant as particle concentration is constant, the net change of particles transported through the cotransporter is different as the number of particles involved in cotransport expands with $V_p$ or $V_c$ increasing. Based on the initial conditions we give and the general conditions in practice, the $\rm H^+$ concentration is much higher in periplasm than in cytoplasm (possibly by an order of magnitude or more), and lactose concentration is generally higher in cytoplasm than in periplasm. In the case of $\xi=0$, the net transport is a 1:1 cotransport for both kinds of particles from periplasm to cytoplasm, so when periplasm volume $V_p$ fixed and cytoplasm volume $V_c$ varied, the net transport does not change much and therefore the time to reach equilibrium does not change significantly. But in contrast the net transport increases almost linearly when $V_c$ fixed and $V_p$ varied, so that the time required to reach equilibrium also increases monotonically with the volume ratio $V_p/V_c$. 

	The above is the case of $\xi=0$. And for the case of $\xi=1$, it is verified that the conclusion about the ratio of equilibrium solutions, $i.e.$ , ${(N_c^S/V_c)}/{(N_p^S/V_p)}=\rm const.$ , holds for both $\rm H^+$ and lactose particles within error permissibility when fixing the number of each particle in the initial condition. Then the next concern is the time required to reach the equilibrium solution when $\xi=1$, and we still require a fixed concentration of the particles in periplasm and cyptoplasm in the initial conditions. The following figures Fig~\ref{5}(a) and Fig~\ref{5}(b) show the relations between $V_p/V_c$ ($V_c$ fixed) or $V_c/V_p$ ($V_p$ fixed), and the time to reach the equilibrium solution, respectively, and we consider that the particle $S$ reaches equilibrium when the rate of change of $N_c^S$ is always less than $10^{-8}$. It is easy to see that in the case of $\xi=1$ there are some phenomena that do not match our expectations. For example, with $V_c$ fixed, the time to reach equilibrium for both kinds of particles increase monotonically with the increase of $V_p$; however, with $V_p$ fixed, the equilibrium time for $\rm H^+$ decreases monotonically with $V_c$ increasing, but the equilibrium time for lactose shows an increase followed by a decrease and takes a maximum around $V_c/V_p=4$. 
\begin{figure}[!h]
\centering
\includegraphics[scale=0.9]{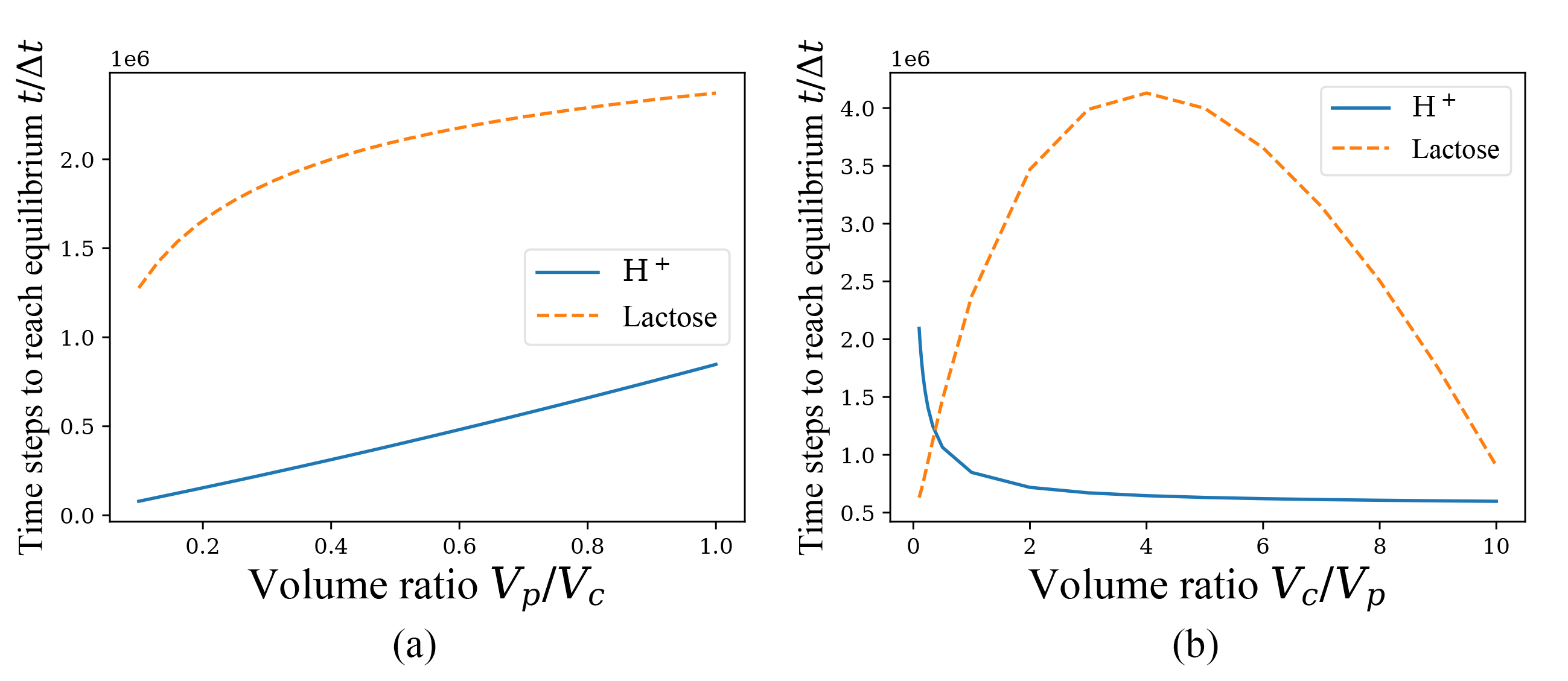}
\caption{{\bf Time to reach equilibrium for $\xi=1$ when $V_c$ or $V_p$ is fixed and the other one varies.}
 Different time for $\rm H^+$ and lactose to reach equilbrium state with the unit of $\Delta t$ when the initial concentrations of $\rm H^+$ and lactose are fixed in periplasm and cytoplasm, which has the initial conditions $n_1=10^3, n_{k\neq1}=0, N_p^{\rm H^+}/V_p=C_0/10, N_c^{\rm H^+}/V_c=C_0/1000, N_p^{\rm L}/V_p=C_0/40, N_c^{\rm L}/V_c=3C_0/40$, $\xi=1$. (a) $V_c$ fixed to $10^6/C_0$. (b) $V_p$ fixed to $10^6/C_0$.}
\label{5}
\end{figure}

	The above two phenomena are quite anomalous, but they can still be broadly explained as follows. Since the concentration of $\rm H^+$ and lactose on both sides of the cell membrane and the number of cotransporters are all constant, the transport rate of the cotransporter changes little when varying volume ratio, so the main difference in the equilibrium time comes from the number of particles by net cotransport. The $\rm H^+$ concentration is considerably higher in periplasm than in cytoplasm, with lactose concentration the opposite. Therefore, according to the proportional relationship that needs to be satisfied by the equilibrium solution, the net transport of $\rm H^+$ is from periplasm to cytoplasm, which eventually makes the $\rm H^+$ concentration in cytoplasm much higher than in periplasm. Then increasing $V_p$ causes the net transport of $\rm H^+$ to increase significantly, so the equilibrium time also increases significantly and monotonically with $V_p/V_c$ up. Meanwhile, the net transport of lactose is from cytoplasm to periplasm, so the effect of changing $V_p$ on the net transport of lactose is positive, and therefore the equilibrium time increases with $V_p/V_c$ up, but asymptotic phenomena occur as $V_p/V_c$ continues to increase. Thus, the increasing rate of the time for lactose to reach equilibrium slows down rapidly as $V_p/V_c$ increases, but is still greater than the $\rm H^+$ equilibrium time.
	
	When varying $V_c$, the case of $\rm H^+$ is similar to the earlier discussion and easily explained as monotonically decreasing with $V_c/V_p$, but the case of lactose is more complicated. Due to the proportionality of the equilibrium solution, when $V_c$ is very large or small, the concentration of lactose in cytoplasm at equilibrium will be close to the initial concentration in cytoplasm or periplasm, respectively, so that the net transport is similar and both are small, but $V_c/V_p$ of moderate size may produce a large net transport. This may lead to a phenomenon that equilibrium time of lactose increases and then decreases with $V_c/V_p$ up as shown in Fig~\ref{5}(b), but the exact situation still needs to be supported by experimental data. From the above discussion we can find that in the case of $\xi\neq0$, the correlation between the equilibrium of the two particles is very weak, not only the equilibrium solution itself is irrelevant, but also the correlation of the time required to reach the equilibrium solution is very low, which is exactly an important property of this model. This leads a slightly questioning of the model when $\xi\neq0$.

	Above we set the volume ratio of periplasm and cytoplasm as the horizontal coordinate to calculate, but for a fixed $E.\ coli$ cell (or cells from the same population), periplasm is the portion between the cell wall (cytoderm) and the cell membrane, and cyptoplasm is the portion within the cell membrane other than the nuclear region, and the sum of the volumes $V_c+V_p$ varies very little due to the rigidity of the cytoderm. Therefore, we continue to investigate its effect on the convergence rate by taking the percentage of periplasm to the sum of periplasm and cyptoplasm volumes as the horizontal coordinate while fixing the sum of the two volumes. In this case, we choose to fix the initial concentration and initial number of particles in periplasm and cyptoplasm, respectively, and calculate the time required for different particles to reach equilibrium. The following Fig~\ref{6} is the image of the time required to reach equilibrium as the periplasm fraction changes for the case $\xi=0$, where it is still considered that the particle $S$ reaches the equilibrium solution when the change rate of $N_c^S$ is always less than $10^{-9}$. As seen from the experimental data in article \cite{stock1977periplasmic}\cite{graham1991periplasmic}, the range of horizontal ordinates in Fig~\ref{6} already includes the volume fraction of periplasm (8\%-40\%) in which $E.\ coli$ can survive.
\begin{figure}[!h]
\centering
\includegraphics[scale=0.9]{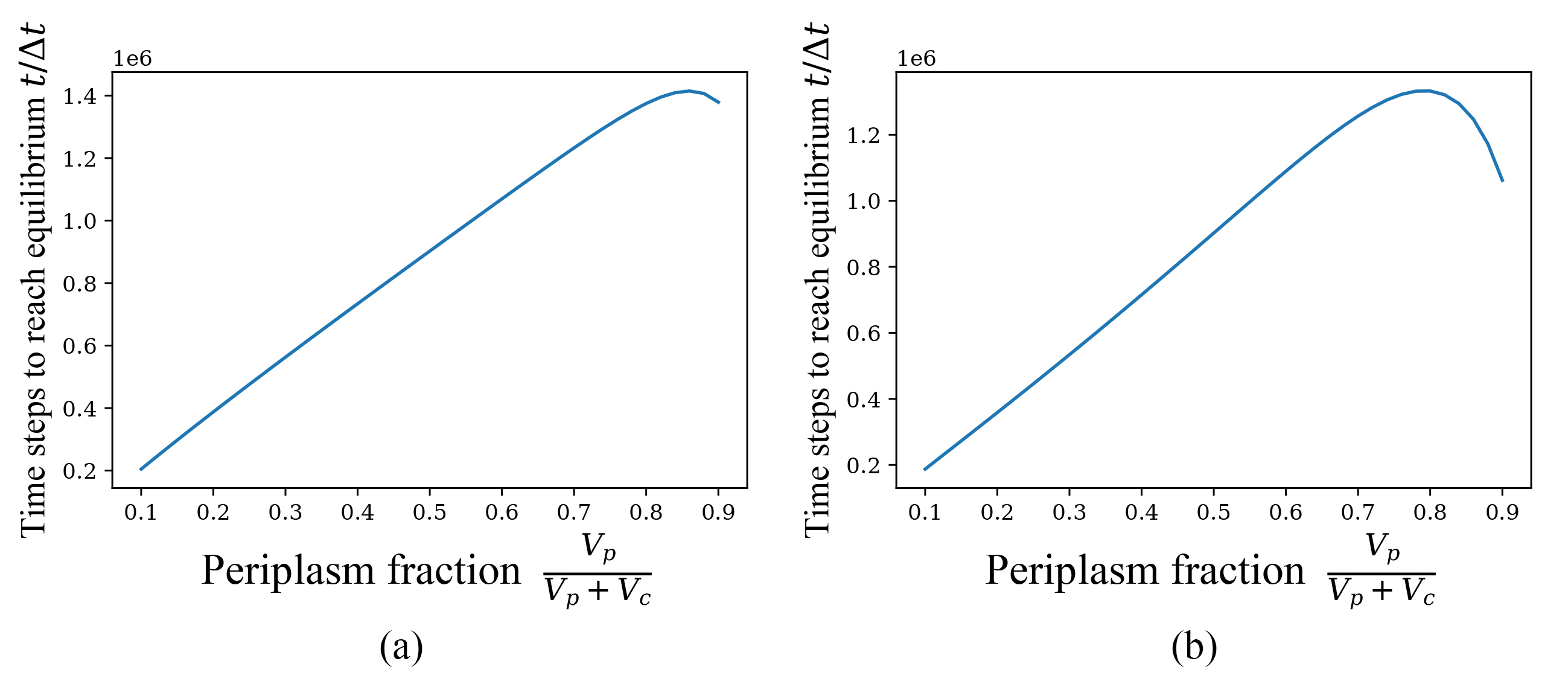}
\caption{{\bf Time to reach equilibrium with the variation of periplasm fraction $\frac{V_p}{V_p+V_c}$, $\xi=0$.}
The sum of volumes of periplasm and cytoplasm is fixed, $V_c+V_p=2\times10^6/C_0$. (a) Initial particle concentration fixed, and the initial conditions are like $n_1=10^3, n_{k\neq1}=0, N_p^{\rm H^+}/V_p=C_0/10, N_c^{\rm H^+}/V_c=C_0/1000, N_p^{\rm L}/V_p=C_0/40, N_c^{\rm L}/V_c=3C_0/40$. (b) Initial particle population fixed, and the initial conditions are like $n_1=10^3, n_{k\neq1}=0, N_p^{\rm H^+} =10^5, N_p^{\rm L}=2.5\times10^4, N_c^{\rm L}=7.5\times10^4, N_c^{\rm H^+}=10^3$.}
\label{6}
\end{figure}

	It can be found that the two graphs have some similarity, both appearing to be approximately linearly increasing followed by monotonically decreasing. Although it may seem strange, the above case of $\xi=0$ can actually be roughly explained by the standard theoretical methods. With a fixed initial concentration, the $\rm H^+$ concentration in cytoplasm increases continuously as the volume fraction of periplasm increases due to the higher concentration of $\rm H^+$ in periplasm than in cyplasm in the initial conditions .However, the increase does not exceed the initial $\rm H^+$ concentration in periplasm, meanwhile the volume of cyptoplasm decreases accordingly. If the volume difference between periplasm and cytoplasm is too large, and the $\rm H^+$ concentrations in the smaller part of the volume at equilibrium will be close to the initial values of the larger part, then the net transport will be neither large, so the net transport of hydrogen ions is likely to increase and then decrease as the volume fraction of periplasm increases. Therefore, the time to reach equilibrium is likely to increase and then decrease at constant initial concentration and constant cotransport rate.

	With a fixed initial particle number, $\left|\Delta G\right|$ decreases as the fraction of periplasm increases, resulting in a decrease in the net amount of particles transported, but the change in volume also decreases the transport rate of cotransporters from periplasm to cytoplasm but increases the rate of reverse direction, with a decrease in the (initial) net transport rate. A qualitative analysis of the form reveals a more pronounced change in the net transport rate (than in the net amount of particles transported), but the change rate of the net transport rate gradually decreases as the fraction of periplasm increases, thus possibly leading to an increase and then a decrease in the time required to reach equilibrium. The above analyses and explanations are qualitative and still need to be verified or denied by experimental data. The above is the case of $\xi=0$, while the following Fig~\ref{7}(a) and Fig~\ref{7}(b) are the case of $\xi=1$, which are surprisingly not similar, unlike the similarity of the previous Fig~\ref{6}(a) and Fig~\ref{6}(b).
\begin{figure}[!h]
\centering
\includegraphics[scale=0.9]{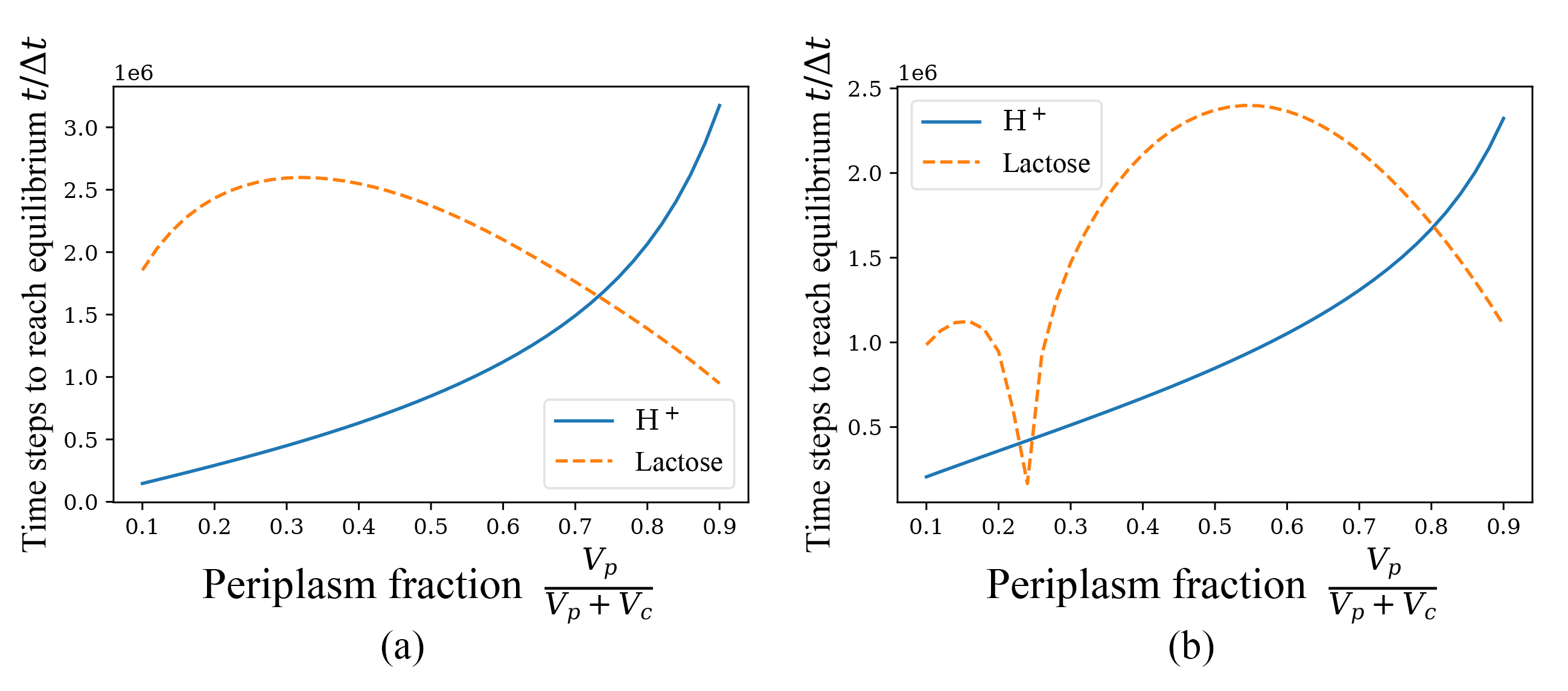}
\caption{{\bf Time to reach equilibrium with the variation of periplasm fraction $\frac{V_p}{V_p+V_c}$, $\xi=1$.}
 Different time for $\rm H^+$ and lactose to reach equilbrium state with the unit of $\Delta t$. The sum of volumes of periplasm and cytoplasm is fixed, $V_c+V_p=2\times10^6/C_0$. (a) Initial particle concentration fixed, and the initial conditions are like $n_1=10^3, n_{k\neq1}=0, N_p^{\rm H^+}/V_p=C_0/10, N_c^{\rm H^+}/V_c=C_0/1000, N_p^{\rm L}/V_p=C_0/40, N_c^{\rm L}/V_c=3C_0/40$. (b) Initial particle population fixed, and the initial conditions are like $n_1=10^3, n_{k\neq1}=0, N_p^{\rm H^+} =10^5, N_p^{\rm L}=2.5\times10^4, N_c^{\rm L}=7.5\times10^4, N_c^{\rm H^+}=10^3$.}
\label{7}
\end{figure}

	A closer observation reveals that although the equilibrium time images of lactose are disparate, the equilibrium time images of $\rm H^+$ are still similar in shape. This is primarily due to the fact that in the presence of leakage, as mentioned earlier, the equilibrium of the two kinds of particles has little influence on each other. And no matter initial concentration or initial particle numbers is fixed, in the range of periplasm volume fraction in our calculation (${V_p}/{(V_p+V_c)}\in[0.1,0.9]$), the net transport direction of hydrogen ions is from periplasm to cytoplasm according to $\Delta G$. Increasing the periplasm volume proportion in the presence of leakage, when the initial concentration is fixed, the (initial) net transport rate remains constant and the net transport amount increases, while when the initial particle number is fixed, the (initial) net transport rate decreases and the net transport volume does not change much from the aforementioned proportionality of the equilibrium solution, so the time required to reach equilibrium in both cases shows a monotonically increasing state. 

	The case of lactose is more complicated, especially in the condition of fixed initial particle number, where there are three inflection points and also the order for $\rm H^+$ and lactose to reach equilibrium changes with different volume fraction of periplasm. We can give the following explanation for the minimal value point generated near ${V_p}/{(V_p+V_c)}=0.25$ under the condition of fixed initial particle numbers. Since the initial value of $N_c^{\rm L}/N_p^{\rm L}={1}/{3}$, when $V_c/V_p$ is around 3, the initial condition of lactose is actually very close to the equilibrium state because the total number of cotransporters is small compared to the number of particles. $N_c^{\rm L}$ and $N_p^{\rm L}$ change very little during the cotransport process and can be considered as reaching equilibrium very early. Changing the initial value of lactose and computationally verifying, it is found that this phenomenon is practically universal. When the volume ratio is adjusted so that the initial value of lactose is close to the equilibrium solution, there is always a significant reduction in the time for lactose to reach equilibrium, causing lactose's reaching equilibrium before $\rm H^+$. More generally, a similar phenomenon occurs for both kinds of particles. Nevertheless, it should still be noted that for the general realistic initial values, as in the previous analysis, $\rm H^+$ usually reaches equilibrium before lactose because of the additional leakage current regulation, which also complements and explains the above inference. Regarding the remaining phenomena embodied in the two figures, it is difficult for the author to give a reasonable explanation here, and I can only leave it to the experimental data to verify or deny.
	\subsection*{Effect of changing the initial concentration and initial distribution of particles on equilibrium properties}
	Now we will explore the effect of the initial state on the equilibrium properties. Inspired by the previous section, we can observe the effect of changing the initial state on $\Delta G$ to determine the effect on the equilibrium solution and the time required to reach equilibrium. We still study only the $\xi=0,1$ cases in the following, as in the previous section. If the number of particles in the initial state is expanded linearly, and $\Delta G$ does not change in this case, then it can be presumed that the equilibrium solution and the time to reach equilibrium should also increase roughly linearly, which is verified to be correct when $\xi=0,1$ after calculation, and we will not elaborate it here. In the following we will consider the effect of changing the initial distribution of particles in periplasm and cytoplasm. Since both volumes $V_c,V_p$ are kept constant, we do not need to care about the difference between the two conditions of initial concentrations and initial particle numbers. The following Fig~\ref{8} shows the change of $\rm H^+$ and lactose equilibrium fractions in periplasm with the variation of $\rm H^+$ initial fraction in periplasm. 
\begin{figure}[!h]
\centering
\includegraphics[scale=1.0]{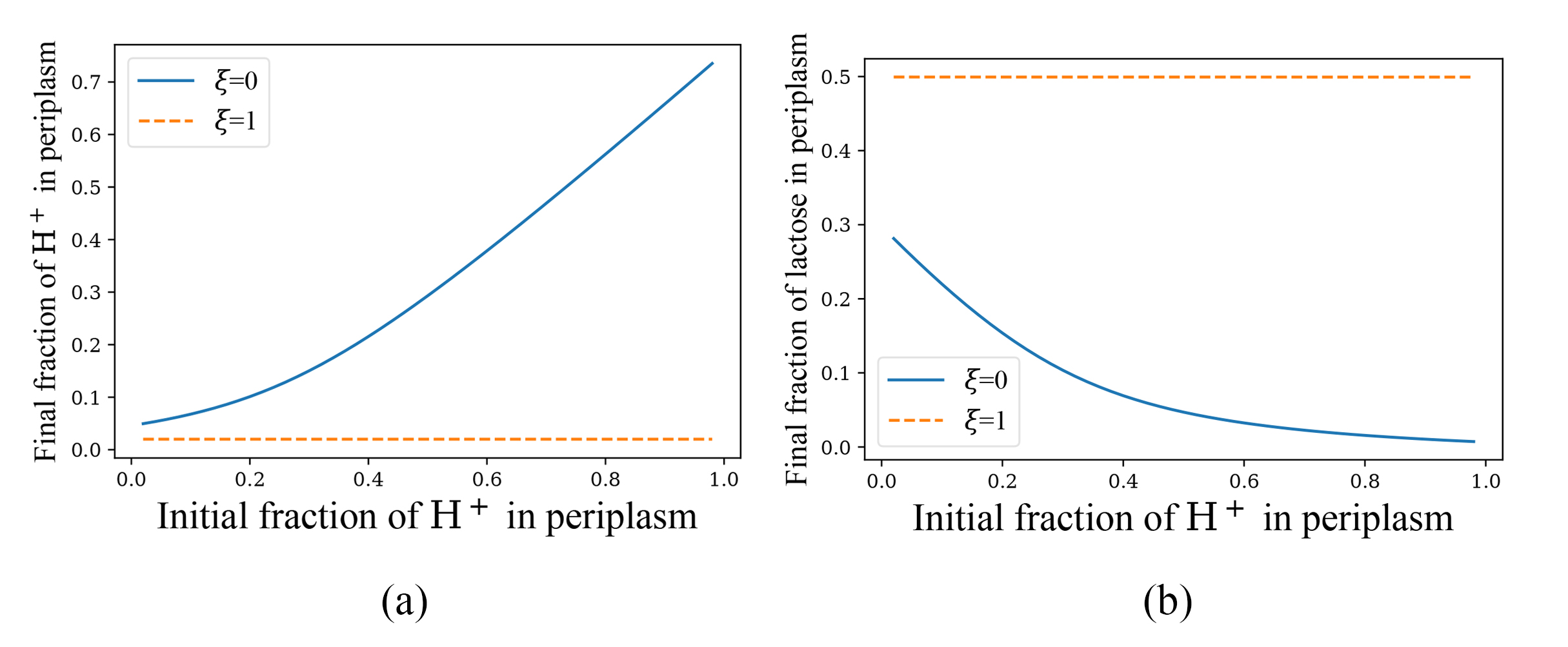}
\caption{{\bf Equilibrium fractions of particles in periplasm as initial fraction of $\rm H^+$ in periplasm changes.}
The initial conditions are $n_1=10^3, n_{k\neq1}=0, N_p^{\rm H^+}+N_c^{\rm H^+}=10^5, N_p^{\rm L}=2.5\times10^4, N_c^{\rm L}=7.5\times10^4$. And also, we set $V_c=V_p=10^6/C_0$, $\xi=0,1$. There is always enough time for particles to stabilize so the final fractions equals the equilibrium fractions. (a) Final $\rm H^+$ fraction. (b) Final lactose fraction.}
\label{8}
\end{figure}

	It can be seen that the case of $\xi=1$ conforms well to our expectation, and the equilibrium solution scale relation hardly changes with the initial conditions. In the case of $\xi=0$, the equilibrium proportion of $\rm H^+$ in periplasm increases monotonically with the initial proportion of $\rm H^+$ in periplasm, and it can be observed that equilibrium value is always lower than the initial value. Meanwhile, the proportion of lactose in periplasm in the equilibrium solution decreases monotonically with the proportion of $\rm H^+$ in periplasm in the initial condition. Both phenomena are consistent with common biological sense. When the $\rm H^+$ concentration in periplasm is higher than in cytoplasm, $\rm H^+$ provides a sufficient electrochemical potential for the transportation of lactose to cytoplasm, which allows $E.\ coli$ to take up lactose against the concentration gradient of it. But doing so depletes $\rm H^+$ in periplasm, making the $\rm H^+$ concentration gap between periplasm and cytoplasm smaller when equilibrium is reached. The larger the initial $\rm H^+$ concentration difference between periplasm and cyptoplasm, the more $\rm H^+$ can be used and the more lactose is transported to cyptoplasm. Also, because of the need to maintain a high concentration of lactose in the cytoplasm, the periplasm must maintain a sufficiently high $\rm H^+$ concentration at equilibrium to counteract the chemical potential of lactose. Due to the fact that some of the $\rm H^+ $ are transported into cytoplasm with lactose, the $\rm H^+$ concentration difference at equilibrium is lower than in the initial condition. The above two phenomena are thus explained. In the following, we consider the change in the initial proportion of lactose in periplasm, corresponding to the change in the proportion of two particles in periplasm at equilibrium, and the figure \ref{11} is shown below. The following Fig~\ref{9} show the variations in $\rm H^+$ and lactose equilibrium fraction as the initial fraction of lactose in periplasm changes.
\begin{figure}[!h]
\centering
\includegraphics[scale=1.0]{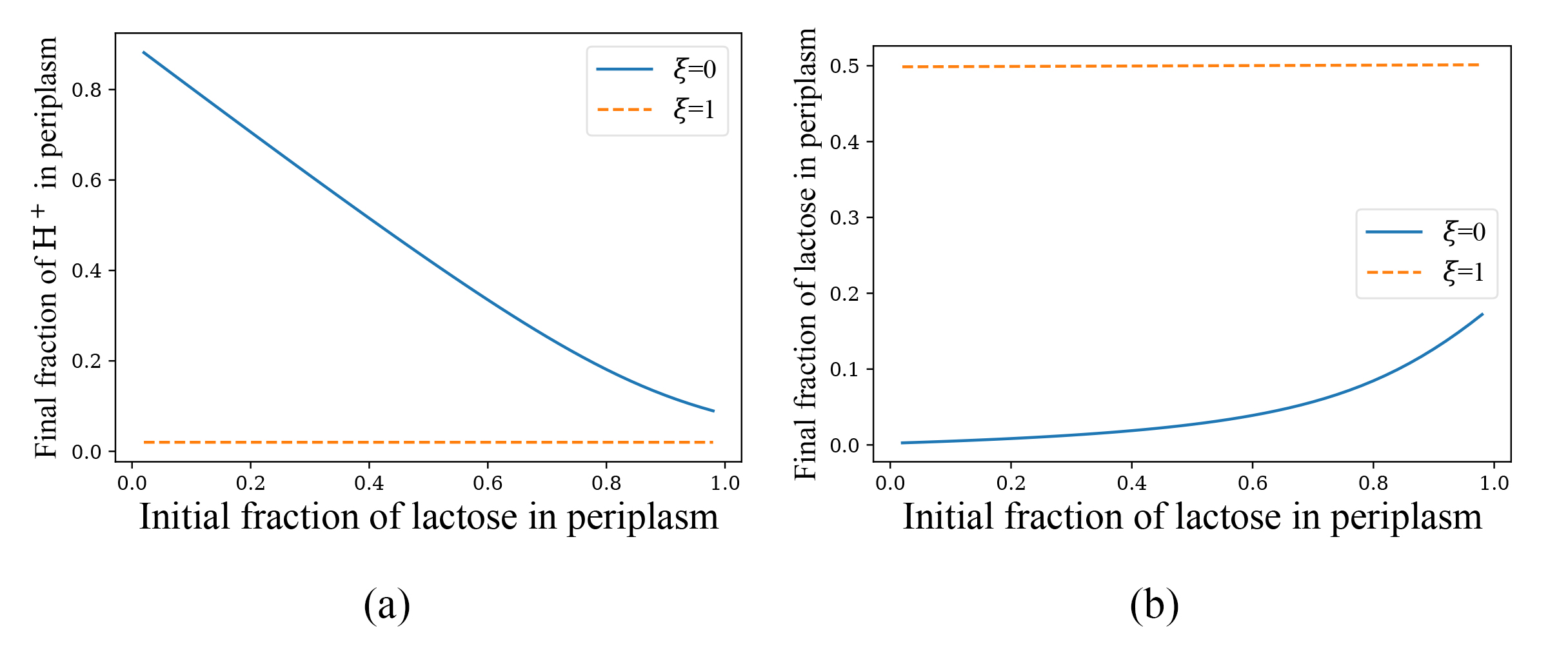}
\caption{{\bf Equilibrium fractions of particles in periplasm as initial fraction of lactose in periplasm changes.}
The initial conditions are $n_1=10^3, n_{k\neq1}=0, N_p^{\rm H^+}=9\times10^4, N_c^{\rm H^+}=10^4, N_p^{\rm L}+N_c^{\rm L}=10^5$. And also, we set $V_c=V_p=10^6/C_0$, $\xi=0,1$. There is always enough time for particles to stabilize so the final fractions equals the equilibrium fractions. (a) Final $\rm H^+$ fraction. (b) Final lactose fraction.}
\label{9}
\end{figure}

	The above two plots obtained by varying the initial distribution of lactose are also fully interpretable in a similar way as above. When $\xi=0$, as the initial lactose molecules in periplasm keep increasing, the absolute value of the free energy change of cotransport in the initial state $\left|\Delta G\right|$ increases, and more lactose molecules are transported into cytoplasm. As the two particles are transported into cytoplasm in strictly equal amounts when $\xi=0$, there will be a monotonic decrease in the amount of hydrogen ions and a slow monotonic increase in the amount of remaining lactose in periplasm at equilibrium with initial fraction of lactose in periplasm increasing. The $\xi=1$ cases still meet the expectation and we do not elaborate on it.
	\subsection*{Effect of changing the total number of cotransporters on equilibrium properties}
\begin{figure}[!h]
\centering
\includegraphics[scale=1.0]{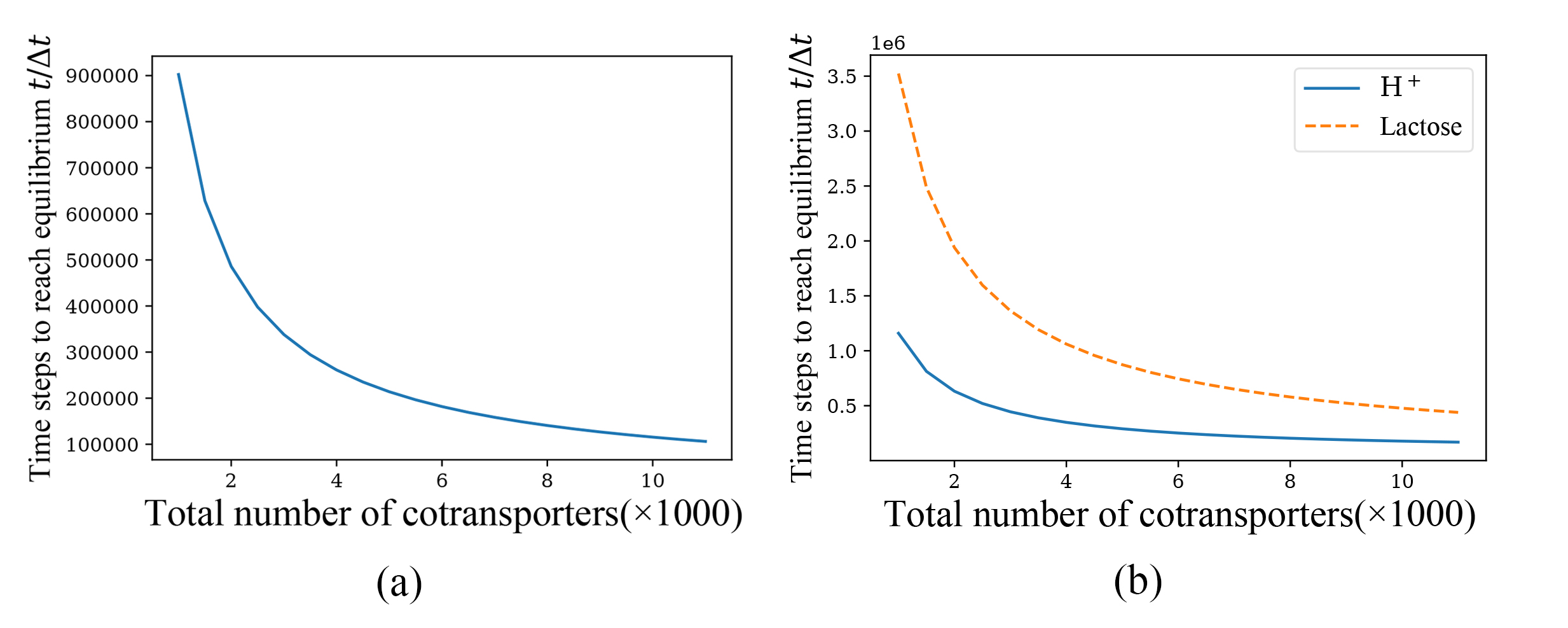}
\caption{{\bf Time to reach equilibium with the variation of the total number of cotransporters.}
The initial conditions are $N_p^{\rm H^+}=10^5, N_c^{\rm H^+}=10^3, N_p^{\rm L}=2.5\times10^4, N_c^{\rm L}=7.5\times10^4, V_c=V_p=10^6/C_0$. All cotransporters are supposed to be in state 1 initially. (a) The $\xi=0$ case. (b) The $\xi=1$ case.}
\label{10}
\end{figure}

	From the master equations of the system, we know that the first order derivatives of both particle numbers and cotransporter numbers in different states are linearly related to the current values of some states of the cotransporters. So we guess that the convergence rate should be linearly related to the total number of cotransporters with constant initial values, and then the time required to reach the equilibrium state should be inversely proportional to cotransporter numbers.  After the calculation we have Fig~\ref{10} for $\xi=0,1$ cases.

	Although the negative correlation is evident in both plots, a closer observation does not show the inverse proportion relationship as we would expect. The reason for this is unclear and needs to be verified or disproved by experimental data.
\section*{Conclusions and further discussions of the above results}
	For the cotransport process of $E.\ coli$ LacY protein symporting $\rm H^+$ and lactose, we apply the random-walk model proposed in article \cite{barreto2020random} to computationally simulate and find, validate or predict the following phenomena. \begin{itemize}
	\item In the absence of leakage, the concentrations of the two particles $\rm H^+$ and lactose reach stability simultaneously; in the presence of leakage, the time required for the two particles to reach the equilibrium state is separated. In the realistic situation where the number of cotranporters is small relative to the number of transported particles, the intensity of leakage, if leakage exists, hardly affects the equilibrium state of cotransport, but affects the time for both particles to reach the equilibrium state, and generally the more pronounced leakage is, the shorter the time to reach equilibrium. 
	\item In the absence of leakage, fixing the initial number or concentration of $\rm H^+$ and lactose in periplasm and cytoplasm, for homogenous $E.\ coli$ cells with different periplasm and cytoplasm volumes, the periplasm volume but not the cytoplasm volume has a greater effect on the equilibrium state and the time required to reach equilibrium. In other words, when periplasm volumes are similar and cytoplasm volumes are different, cells have similar equilibrium states and need similar time to reach equilibrium by this cotransport process, but not vice versa. 
	\item In the presence of leakage, fixing the initial concentration of $\rm H^+$ and lactose in periplasm and cytoplasm, for homogenous $E.\ coli$ cells with different periplasm and cytoplasm volumes, the time required for the pH of cytoplasm to stabilize increases monotonically with the periplasm to cytoplasm volume ratio increasing. Meanwhile,  the time for the lactose concentration to reach equilibrium has a more complex relationship with the volume ratio of periplasm and cytoplasm as shown in Fig~\ref{5}.
	\item For a certain $E.\ coli$ cell (or cells with similar size from the same population), fixing the initial number or concentration of $\rm H^+$ and lactose in periplasm and cytoplasm, the time for the pH of cytoplasm to stabilize increases monotonically as the cell loses water regardless of the presence or absence of leakage (under the condition of cell survival), whereas the time for the concentration of lactose to stabilize varies more complexly, as seen in Fig~\ref{6} and Fig~\ref{7}.
	\item For $E.\ coli$ cells from the same population, the concentration ratio of particles in periplasm and cyptoplasm in equilibrium state does not vary with the initial state concentration ratio if leakage exists.
	\item For different subpecies of $E.\ coli$, the time for the cotransport process to reach equilibrium is negatively but not inversely correlated with the amount of the cotransporter LacY with initial concentrations of $\rm H^+$ and lactose in periplasm and cytoplasm fixed.
\end{itemize}
	Many of the above phenomena can be qualitatively explained, but some of them still can not be well explained and need to be verified or negated by experimental data.

	After summarizing the results above, it is easy to notice that in the case of the parameter $\xi=0$, $i.e.$ , no leakage, the results are mostly consistent with biological or chemical intuition, but the appearance of some properties for $\xi\neq0$ case is quite anomalous. The biggest problem is the equilibrium solution. When $\xi\neq0$, the equilibrium solution almost strictly satisfies the concentration proportionality under the general condition that the number of cotransporters is much smaller than the number of particles to be transported. For the cotransport of $E.\ coli$ LacY protein, the concentration of lactose in periplasm and cytoplasm at equilibrium is almost equal because the lactose molecule is neutral, and this relationship does not change with other conditions such as the initial value. This phenomenon may be somewhat different from the reality. Actually for this model in $\xi\neq0$ cases, any neutral particles involved in cotransport has equal concentration in periplasm and cytoplasm, which is no different from ordinary diffusion and far from the energy-consuming active transport. Next, from a kinetic point of view, there are still some parts of the model that need discussions when the parameter $\xi\neq0$. Although the essence of the model is cotransport, we find that the correlation of the time required for two particles to reach equilibrium is not strong. If the initial state of a certain particle happens to satisfy the proportionality relation that the equilibrium state should satisfy, it will soon equilibrate and is hardly controlled by the other particle. From this point of view, the two kinds of particles are almost independent of each other. At this point, the effect of  on lactose transport is more like a facilitated transport, which is also used to explain the uncoupled cotransport in recent years, such as literature \cite{kaback2015chemiosmotic} and \cite{
kaback2019takes}. In contrast, there is no case where some particle reaches equilibrium by itself when $\xi=0$. Therefore, although changing the parameter $\xi$ gives the model good kinetic properties, such as adjusting $\xi$ to control the speed of convergence, the model still has some defects. The parameter $\xi$ is introduced to solve the problem of leakage current, but this solution is not perfect, and we may still need to think about other solutions. And furthermore, actually there is no experimental evidence showing this uncoupled sugar translocation (``leakage") could happen before the concentrations of lactose across the cytoplasmic membrane reach equilibrium. In other words, all the above calculations and explanations are based on a shared assumption, and the specific mechanism of cotransporter LacY still lacks knowledge.
\section*{Acknowledgement}
	I would like to give my sincere gratitude to Prof. Yunxin Zhang from the School of Mathematical Sciences, Fudan University. From my freshman year in Fudan, I have been in contact with Professor Zhang, who is although not my tutor but still gives me much selfless guidance and help in my academic learning and undergraduate studies. It was with his advice and help that I was able to complete this paper as an undergraduate. He taught me almost all the basics about academic writing, read my drafts and made many valuable suggestions. Once again, my sincere thanks to all the people who helped me during the completion of this paper.

%
%
%

\end{document}